\documentclass[aps,pra,twocolumn,showpacs]{revtex4}   

\begin{document}   
\title{Optimum unambiguous discrimination   
of two mixed quantum states}

\author{Ulrike Herzog$^{1}$}   
\author{J\'anos A. Bergou$^{2}$}   
\affiliation{$^{1}$Institut f\"ur Physik, Humboldt-Universit\"at    
Berlin, Newtonstrasse 15, D-12489 Berlin, Germany}   
\affiliation{$^{2}$Department of Physics, Hunter College, City   
University of New York, 695 Park Avenue, New York, NY 10021,   
USA}   
\begin{abstract}   
We investigate generalized measurements, based on positive-operator-valued measures, 
and von Neumann measurements 
for the unambiguous discrimination of two mixed quantum states that occur
with given prior probabilities. 
In particular, we derive the conditions under which the failure  
probability of the measurement can reach its absolute lower bound, 
proportional to the fidelity of the states. The optimum measurement strategy 
yielding the fidelity bound of the failure probability  
is explicitly determined for a number of cases. One example involves two density 
operators of rank $d$ that jointly span a $2d$-dimensional Hilbert space and are 
related in a special way. We also present an application of the results to the problem of unambiguous quantum state comparison, generalizing the optimum strategy for arbitrary prior probabilities of the states.     
\end{abstract}   
\pacs{PACS:03.67.Hk,03.65.Ta,42.50.-p}   
\maketitle   
  
Many applications in quantum communication and quantum cryptography 
are based on transmitting quantum systems that, with given prior probabilities, 
are prepared in one from a set of known mutually nonorthogonal states.   
Since perfect discrimination between nonorthogonal quantum states is impossible, 
measurement strategies for state discrimination have been developed 
that are optimized with respect to various criteria \cite{springer}.  
Here we consider unambiguous discrimination,
requiring that the outcome of the measurement be error-free.
For two mixed quantum states unambiguous discrimination is possible with a finite probability of success if the supports \cite{support} of their density operators are not identical. When the measurement fails, it returns an inconclusive answer but never an error. In the optimal measurement strategy the failure probability is minimum.  
 
The problem of unambiguously discriminating mixed quantum states arises for instance when    
given pure states undergo a specified decoherence process during transmission  
through a quantum channel, or when the quantum system is known to be in a pure 
state which has to be assigned  to a particular set out of a number of given sets 
of pure states, each set corresponding to a mixed state.  
While for two pure states the minimum failure probability  
has been known since long \cite{ivan,jaeger},  
the study of unambiguous discrimination among mixed states,  
or sets of pure states, respectively, started  
only recently \cite{SBH,BHH,rudolph,raynal,feng1,eldar,HB2}.  
A complete solution, determining the minimum achievable failure probability for 
arbitrary prior probabilities of the states, has been obtained for the special 
cases of discriminating a pure and a mixed state \cite{SBH,BHH}, 
and of two mixed states of rank $d$ 
in a $(d+1)$-dimensional joint Hilbert space \cite{rudolph}.  
For discriminating two arbitrary mixed states,   
bounds have been derived for the failure probability \cite{rudolph}, 
in terms of the fidelity of the states. 
In this paper we perform a more detailed analysis, investigating the  
conditions under which the lowest bound, proportional to the fidelity,  
can be reached, and deriving also the von Neumann measurements for unambiguous discrimination.  
 
We start by recalling that a measurement for distinguishing two quantum states, 
characterized by the density operators $\rho_1$ and $\rho_2$ 
and the prior probabilities $\eta_1$ and $\eta_2= 1-\eta_1$, respectively,     
can be formally described by three positive operators $\Pi_k$  
with $\sum_{k=0}^{2} \Pi_k =I$, 
where $I$ is the identity. These detection  
operators are defined in such a way that  
$\rm Tr(\rho\Pi_k)$ 
 with $k=1,2$ is the probability that a system   
prepared in a state $\rho$ is inferred to be in the state $\rho_k$,  
while $\rm Tr(\rho\Pi_0)$ is the probability that the     
measurement fails to give a definite answer. 
When all detection   
operators are projectors, the measurement is a von Neumann measurement,  
otherwise it is  
a generalized measurement based on a positive operator-valued measure (POVM).  
From the detection operators $\Pi_k$  
schemes for realizing the measurement can be obtained  
\cite{neumark,preskill}.        
 
It is our aim to investigate the optimum measurement strategy  that minimizes 
 the total failure probability  
\begin{equation} 
Q= \eta_1 \rm Tr (\rho_1\Pi_0) + \eta_2 \rm Tr (\rho_2\Pi_0).  
\end{equation} 
From the relation between the arithmetic and the geometric mean  
and from the  
Cauchy-Schwarz-inequality \cite{nielsen,feng1}   
it follows that  
$Q     \geq   2\sqrt{\eta_1 \eta_2 
           \rm Tr(\rho_1\Pi_0)\rm Tr(\rho_2\Pi_0)}  
  \geq  2\sqrt{\eta_1 \eta_2}\,  
 {\rm Max}_U\, |{\rm Tr}(U\sqrt{\rho_1}\Pi_0\sqrt{\rho_2})|$, 
where $U$ describes an arbitrary unitary transformation. 
The failure probability takes its absolute minimum when the  
two equality signs hold. This is true if and only if both the relations  
$\eta_1 \rm Tr(\rho_1\Pi_0)= \eta_2 \rm Tr(\rho_2\Pi_0)$ 
and $U \sqrt{\rho_1}\sqrt{\Pi_0} \sim \sqrt{\rho_2}\sqrt{\Pi_0}$ are fulfilled.  
From the first relation we conclude that 
the number of    
inconclusive results is equally distributed  
among the two incoming states. After multiplying the second relation  
with its Hermitean  
conjugate, the two conditions for equality can be combined to yield 
$\sqrt{\Pi_0}(\eta_2 \rho_2 - \eta_1 \rho_1)\sqrt{\Pi_0}  =  0$. 
Since in the POVM-formalism  
the detection operators transform a quantum state according to   
$\rho\rightarrow \sum_k \sqrt{\Pi_k}\,\rho\,\sqrt{\Pi_k}$ \cite{preskill},     
it follows that the total failure probability is smallest when  
in case of failure the two  
density operators  are transformed into states that are identical  
after normalization and therefore cannot be  
further discriminated.   
 
We now recall that unambiguous discrimination of two  
states leads to the requirement  
$\rho_1 \Pi_2 = \rho_2 \Pi_1= 0$ \cite{springer}.   
Substituting  $\Pi_0=I - \Pi_1 - \Pi_2$ into the inequality for the 
failure probability $Q$ \cite{feng1}, given above, we arrive at        
\begin{equation} 
\label{Q2}    
Q   \geq  2\sqrt{\eta_1 \eta_2}  
 {\rm Max}_U\,  
 |{\rm Tr}(U\sqrt{\rho_1}\sqrt{\rho_2})|   
  =  2\sqrt{\eta_1 \eta_2}\; F,  
\end{equation} 
where $F={\rm Tr}\,[\left(\sqrt{{\rho}_2}\; {\rho}_1  
\sqrt{{\rho}_2}\right)^{1/2}]$ is the fidelity \cite{nielsen}.  
Using a different method, it has been found already previously 
by Rudolph et al. \cite{rudolph} that 
\begin{equation}
\label{rud}
Q\geq \left \{ \begin{array}{ll}
2\sqrt{\eta_1 \eta_2}F = Q_0\;\; & \mbox{if 
$ F \leq \sqrt{\frac{\eta_{1}}{\eta_{2}}}\leq \frac{1}{F}$} \\ 
\eta_{\rm min} + \eta_{\rm max}F^2\;\;  & \mbox{otherwise,} 
 \end{array}
\right.
\end{equation}    
with $\eta_{\rm min}$ $(\eta_{\rm max})$ denoting the smaller (larger) of  
the prior probabilities.   
Here, in addition, we obtained the {\it necessary and sufficient conditions}  
that the detection operators have to fulfill in order to reach  
the fidelity bound $Q_0$. They can be summarized as
\begin{eqnarray}   
\label{ns4}   
&&\Pi_0 = I - \Pi_1 - \Pi_2 \geq 0, \;\;\Pi_1\geq 0, \;\;\Pi_2\geq 0,\\    
\label{ns1}   
&&\rho_1 \Pi_2 = \rho_2 \Pi_1= 0,\\     
\label{ns2}   
&&\eta_1 {\rm Tr}(\rho_1\Pi_0)= \eta_1 [1- {\rm Tr}(\rho_1\Pi_1)]      
    =\sqrt{\eta_1 \eta_2}\,F,\\     
\label{ns3}   
&&\eta_2 {\rm Tr}(\rho_2\Pi_0)= \eta_2 [1- {\rm Tr}(\rho_2\Pi_2)]      
    =\sqrt{\eta_1 \eta_2}\,F. 
\end{eqnarray}   
 
In the following we investigate the conditions  
under which detection operators exist that satisfy  
Eqs. (\ref{ns4}) - (\ref{ns3}). For this purpose we use the spectral representations       
\begin{equation} 
\label{rho} 
\rho_1 = \sum_{l=1}^{d_1} r_l |r_l\rangle\langle r_l|, 
\qquad    
\rho_2 = \sum_{m=1}^{d_2} s_m |s_m\rangle\langle s_m|,                  
\end{equation}   
where $r_l, s_m\neq 0$, and $\langle r_l|r_{m}\rangle = \delta_{l,m}   
=\langle s_l|s_m\rangle$. 
Furthermore, we   
introduce the projection operators    
\begin{equation}   
\label{10}   
P_1 = \sum_{l=1}^{d_1} |r_l\rangle\langle r_l|,    
\qquad   
P_2 = \sum_{m=1}^{d_2} |s_m\rangle\langle s_m|,\;\;\;                  
\end{equation}   
and the nonnormalized states    
$|r_l^{\parallel}\rangle= P_2 |r_l\rangle$. We can construct a  
complete orthonormal basis $\{|h_k\rangle \}$    
in the subspace $\mathcal{H}_{1\parallel}$     
spanned by the state vectors $P_2 |r_l\rangle$,    
using the recursion relation   
$|\tilde{h}_k\rangle = P_2|r_k\rangle - \sum_{i=1}^{k-1}     
 |h_i\rangle \langle h_i|P_2|r_k\rangle$ and    
determining  $|h_k\rangle = |\tilde{h}_k\rangle /   
 \|\tilde{h}_k \|$ \cite{nielsen}.        
The dimensionality ${d_{1\parallel}}$ of    
$\mathcal{H}_{1\parallel}$ is equal to the rank    
of the matrix formed by the elements $\langle r_l |P_2| r_n \rangle$.   
Similarly, in the subspace $\mathcal{H}_{1\perp}$ that is  spanned   
by the  nonnormalized vectors $|r_l^{\perp}\rangle=(I-P_2)|r_l\rangle$,   
we can obtain an orthonormal  basis $\{|v_i\rangle \}$  of   
dimension $d_{1\perp}$.  
The respective projection operators into the    
two orthogonal subspaces are    
\begin{equation}   
\label{12}   
P_{1\parallel} = \sum_{k=1}^{d_{1\parallel}}| h_k\rangle\langle h_k|,    
\qquad   
P_{1\perp} = \sum_{i=1}^{d_{1\perp}} |v_i\rangle\langle v_i|,\;\;\;                  
\end{equation}   
where $\rho_2 |v_i\rangle = 0.$  
The operator $P_{10} = P_{1\parallel}+ P_{1\perp}$     
projects onto a subspace $\mathcal{H}_{10}$ of dimension    
$d_{1\parallel} + d_{1\perp}$.     
Noticing that ${\rm Tr}[(P_{10}- P_1)\rho_1]=0$,   
we construct the operator 
\begin{equation} 
\bar{P}_1=  
P_{1\parallel}+ P_{1\perp}-P_1=   
  \sum_{j=0}^{\bar{d}_1}   
  |\bar{r}_j\rangle \langle \bar{r}_j|,  
\label{barP} 
\end{equation} 
where $\rho_1 |\bar{r}_j \rangle = 0$. 
The states $\{|\bar{r}_j\rangle\}$ form an  
orthonormal basis in the 
$\bar{d}_1$-dimensional  
subspace of $\mathcal{H}_{10}$ that is spanned  
by all states    
that are orthogonal to $P_1$, where  
$\bar{d}_1= d_{1\parallel}+d_{1\perp}-d_1$.  
The identity is then given by  
\begin{equation}   
\label{unit1}   
I =   P_{1\perp} + P_2 =  P_{1\perp}+ P_{1\parallel} +  P^{\prime}_2   
= P_1 + \bar{P}_1  +  P^{\prime}_2. 
 \end{equation}   
Here the operator  
$P^{\prime}_2 =  I - P_{1\perp} - P_{1\parallel}$    
projects onto the subspace    
$\mathcal{H}_2^{\prime}$    
spanned by those states     
that are orthogonal to both $P_{1\perp}$ and  $P_{1\parallel}$, implying    
that $\rho_1 P_2^{\prime}=0$.   
Instead of decomposing the eigenstates of $\rho_1$,   
we might as well have started from   
$|s_m\rangle = P_1 |s_m\rangle + |s_m^{\perp}\rangle$,   
obtaining  instead of Eq. (\ref{unit1}) the  
alternative decomposition   
\begin{equation}   
\label{unit2}   
I =   P_{2\perp} + P_1 =  P_{2\perp}+ P_{2\parallel} +  P^{\prime}_1   
= P_2 + \bar{P}_2  +  P^{\prime}_1,   
\end{equation}   
where the projectors are defined analogously. 
 
Now we can specify the general structure of    
all detection operators, $\Pi_1$ and $\Pi_2$, that   
describe unambiguous discrimination, i. e. 
satisfy Eqs. (\ref{ns4}) and (\ref{ns1}). We write   
\begin{equation}   
\label{P1}   
\Pi_1    
= \sum_{j=1}^{d_{1\perp}} \alpha^{\prime}_{j}|v^{\prime}_j\rangle   
\langle v^{\prime}_j|               
=\sum_{i,j=1}^{d_{1\perp}} \alpha_{ij}|v_i\rangle\langle v_j|,   
\end{equation}   
where $0 \leq \alpha^{\prime}_{j}\leq 1$ and  
$|v_j^{\prime}\rangle = \sum_i u_{ji} |v_i\rangle$ with    
$\{u_{ji}\}$ being  a unitary matrix. 
We note that $ \sum_{j} |v^{\prime}_j\rangle   
\langle v^{\prime}_j| = P_{1\perp} $     
since the eigenstates    
$|v_j^{\prime}\rangle$ form a complete orthonormal basis in  
${\mathcal H}_{1\perp}$.     
For representing $\Pi_2$ we start from the same decomposition of  
the identity, and take into account that none of the    
eigenstates of $\Pi_0$ must be contained    
in the subspace $\mathcal{H}_2^{\prime}$ when the failure  
probability is to be as small as possible.  
This leads to    
\begin{equation}   
\label{P2}   
\Pi_2 = \sum_{i=1}^{\bar{d_1}}\beta^{\prime}_{i}|{\bar{r}}_i^{\prime}\rangle   
 \langle {\bar{r}}_i^{\prime}|+ P_2^{\prime}    
 = \sum_{i,j=1}^{\bar{d_1}}    
\beta_{ij}|\bar{r}_i\rangle\langle \bar{r}_j| 
 + I-P_{10},                 
\end{equation}   
where $0 \leq \beta^{\prime}_{i}\leq 1$   
and $\sum_{i=1}^{\bar{d_1}}|\bar{r}^{\prime}_i\rangle   
 \langle \bar{r}_i^{\prime}|= \bar{P}_1$.               
The constants $\alpha_{ij}$ and $\beta_{ij}$ are subject to the   
constraint that $\Pi_0 \geq 0$.    
 
Clearly, when    

$P_1=P_{1\parallel}=I$, and consequently also  
$P_2=P_{2\parallel}=I$, it follows that $\Pi_1=\Pi_2=0$ 
and $\Pi_0 = I$, yielding a unit failure probability that makes 
error-free discrimination impossible. 
We therefore require that   
$P_{1\perp}\neq0$, or $P_{2\perp}\neq0$,  
respectively, 
which, because of normalization, is equivalent to      
\begin{equation} 
\label{norm} 
{\rm Tr}(P_1\rho_2)< {\rm Tr}(P_{1\parallel}\rho_2), 
\quad 
{\rm Tr}(P_2\rho_1)< {\rm Tr}(P_{2\parallel}\rho_1).   
\end{equation} 
 
Before studying the optimum measurement, let us consider  
the von Neumann measurements for unambiguous discrimination.  
If $\alpha_j^{\prime}=0$  for all $j$, and $\beta_i^{\prime} =1$ for all  
$i$, it follows that  
$\Pi_1=0$ and  $\Pi_2=\bar{P}_1+ P_2^{\prime}$. Hence $\Pi_0= P_1$, 
with the failure probability $Q_{N1}  =  \eta_1 + \eta_2{\rm Tr}(P_1\rho_2)$.   
Another von Neumann measurement is generated when     
$\alpha_j^{\prime}=1$  for all $j$, and $\beta_i^{\prime} =0$ for all $i$,   
giving $\Pi_1 = P_{1\perp}$ and $\Pi_2= P_2^{\prime}$.   
Then $\Pi_0 = P_{1\parallel}$, with the failure probability 
\begin{equation}   
\label{QN1p}   
Q_{N1\parallel}  =  \eta_1{\rm Tr}(P_2 \rho_1) +    
\eta_2 {\rm Tr}(P_{1\parallel}\rho_2), 
\end{equation}   
where the relation ${\rm Tr}(P_{1\parallel}\rho_1)   
=1- {\rm Tr}(P_{1\perp}\rho_1)    
={\rm Tr}(P_2 \rho_1)$ has been applied.   
In this measurement the state is unambiguously found to be $\rho_1$   
when a detector click occurs in a direction orthogonal to all eigenstates        
of $\rho_2$. On the other hand, for a click in a direction orthogonal to     
both $P_{1\parallel}$ and $P_{1\perp}$, the state is  
determined to be $\rho_2$ with certainty,    
and in the rest of cases the result is inconclusive.   
So far we relied on Eq. (\ref{unit1}).  
Based on the complementary decomposition of the identity, Eq. (\ref{unit2}),  
we obtain an alternative pair  
of von Neumann measurements. These yield the failure probabilities   
$Q_{N2}  =  \eta_2 + \eta_1{\rm Tr}(P_2\rho_1)$ and    
\begin{eqnarray}   
\label{QN2p}   
Q_{N2\parallel} & = & \eta_2{\rm Tr}(P_1 \rho_2) +    
\eta_1 {\rm Tr}(P_{2\parallel}\rho_1) \ .     
\end{eqnarray}   
Obviously $Q_{N2\parallel} \leq Q_{N1}$ and $Q_{N1\parallel} \leq Q_{N2}$.

We now return to the optimum  measurement.   
Since the von Neumann measurements can be performed 
for arbitrary given parameters, the optimized failure probability certainly  
obeys the inequality 
\begin{equation} 
\label{bar} 
Q_{\rm opt} \leq {\rm Min}\{Q_{N1\parallel},Q_{N2\parallel}\}.   
\end{equation} 
According to Eqs. (\ref{ns2}) and (\ref{ns3}) the absolute minimum  
of the failure probability, $Q_0= 2 \sqrt{\eta_1 \eta_2} \,F,$ is reached  
if and only if the two conditions 
${\rm Tr}(\rho_1\Pi_0)/{F}=\sqrt{{\eta_2}/{\eta_1}}$ and  
${F}/{\rm Tr}(\rho_2\Pi_0)=\sqrt{{\eta_2}/{\eta_1}}$  
are  fulfilled.  
However, due to the structure of the operators $\Pi_1$ and $\Pi_2$,  
the possible values of    
${\rm Tr}(\rho_k\Pi_0)=1-{\rm Tr}(\rho_k\Pi_k)$, for $k=1,2$,    
have a lower bound. 
In particular,   
\begin{eqnarray} 
\label{ineq1} 
{\rm Tr}(\rho_1\Pi_0) & \geq &  1-{\rm Tr}(P_{1\perp}\rho_1) =  
{\rm Tr}(P_2\rho_1), \\    
\label{ineq2} 
{\rm Tr}(\rho_2\Pi_0) & \geq & {\rm Tr}(P_{1\parallel}\rho_2)      
-{\rm Tr}(\bar{P}_1\rho_2)= {\rm Tr}(P_1\rho_2),    
\end{eqnarray} 
where in the first equation the equality sign holds       
when $\alpha_j^{\prime}=1$ in Eq. (\ref{P1}),     
and  in the second equation the equality is reached when  
$\beta_i^{\prime} =1$ in Eq. (\ref{P2}).   
Therefore we obtain that the {\it condition},  
\begin{equation}
\frac{{\rm Tr}(P_2 \rho_1)}{F} \leq \sqrt{\frac{\eta_2}{\eta_1}}     
         \leq \frac{F}{{\rm Tr}(P_1 \rho_2)} \ ,
\label{cond}
\end{equation}
is {\it necessary}, i. e. the fidelity bound, $Q = Q_{0}$, can only be 
reached in part or in the whole of this interval.

The interval specified by Eq. (\ref{cond}) is not empty only when    
${\rm Tr}(P_2 \rho_1){\rm Tr}(P_1 \rho_2) \leq F^2$. 
For two density operators that violate this inequality, 
the failure probability $Q_0$ cannot be achieved for any values of the prior  
probabilities of the states, and the conditions (\ref{ns2}) and  
(\ref{ns3}) are then of no help for determining the optimum measurement.  
 Moreover, our result shows that in general the lower bound $Q_0$ 
can only be reached in an interval of the ratio $\eta_2/\eta_1$ that is smaller 
than the interval  
given in Eq. (\ref{rud}), since ${\rm Tr}(P_2 \rho_1)/{F}\geq F$ 
and  ${F}/{\rm Tr}(P_1 \rho_2)\leq 1/{F}$. 
The latter relations follow from the general inequalities  
\begin{equation}   
\label{aux}   
{\rm Tr}(P_2 \rho_1){\rm Tr}(P_{1\parallel}\rho_2) \geq F^2, 
\quad 
{\rm Tr}(P_1 \rho_2){\rm Tr}(P_{2\parallel}\rho_1) \geq F^2 
\nonumber
\end{equation}   
that can be readily inferred from  Eqs. (\ref{QN1p}), (\ref{QN2p}) and (\ref{Q2}).   
 
The parameter intervals in Eqs. (\ref{rud}) and (\ref{cond}) 
coincide when   
 ${\rm Tr}(P_{1}\rho_2)={\rm Tr}(P_{2}\rho_1)=F^2.$   
This condition is fulfilled, e. g. for density operators 
of the form $\rho_1 = \sum_{i=1}^{d} r_i|r_i\rangle\langle r_i|$ 
and    
$\rho_2 =  \sum_{i=1}^{d} r_i|s_i\rangle\langle s_i|$, 
with 
$\langle r_i| s_j \rangle=b\, \delta_{ij}$,
where the corresponding eigenvalues are identical.  
The fidelity is then  found to be $F=|b|$. 

Another simplification arises when $P_{1\parallel}$ and 
$P_{2\parallel}$ are one-dimensional projectors, 
$d_{1\parallel}=d_{2\parallel}=1$. In this case equality  
holds in Eqs. (\ref{aux}) \cite{footnote} which implies that  
$F^2={\rm Tr}(P_2 \rho_1){\rm Tr}(P_{1\parallel}\rho_2) 
\geq {\rm Tr}(P_2 \rho_1){\rm Tr}(P_1 \rho_2)$, where 
Eq. (\ref{norm}) has been taken into account.   
Hence again for any two density operators  
the necessary condition (\ref{cond}) 
is fulfilled for a certain range of the   
ratio $\eta_2/\eta_1$. 
At the lower limit  
of this range, i. e. for $\sqrt{\frac{\eta_2}{\eta_1}}     
         = \frac{{\rm Tr}(P_1 \rho_2)}{F}$, 
we can write 
$2\sqrt{\eta_1\eta_2}F =  
\eta_1 \frac{F^2}{{\rm Tr}(P_1\rho_2)}+ \eta_2 {\rm Tr}(P_1\rho_2)  
=Q_{N1\parallel}$, and 
similarly we find that at the upper limit 
$2\sqrt{\eta_1\eta_2}F  = Q_{N2\parallel}$.
Thus, if $Q=Q_0$ in the entire range in 
Eq. (\ref{cond}), 
the complete solution for the optimum measurement is known. 

In general, in order to find the optimum measurement strategy that yields
the failure probability $Q_0$, we have to determine the parameters 
$\alpha_{ij}$ and $\beta_{ij}$ in Eqs. (\ref{P1}) and (\ref{P2}) 
that satisfy the necessary and sufficient conditions (\ref{ns4}) - (\ref{ns3}).  
In the following we apply this method to a number of special cases.
  
First we consider two density operators of rank $d$ 
in a $2d$-dimensional joint Hilbert space.
In such a case $P_2^{\prime}=0$ and                   
the identity can be alternatively expressed as  
$I=P_1 + \bar{P}_1$ or $I= P_{1\perp} + P_2$ which means that   
 $P_{1\parallel}=P_2$, $P_{2\parallel}=P_1$ and $\bar{P}_1=P_{2\perp}$. 
We start from  Eqs. (\ref{rho})  with $d_1=d_2=d$ and assume that
$|s_i\rangle = (|r_i\rangle + |{\bar{r}}_i\rangle)/ {\sqrt{2}}$,
and
$|v_i\rangle = (|r_i\rangle - |\bar{r}_i\rangle)/{\sqrt{2}}$ 
($i=1,\ldots,d$).
Then we obtain  $F= \sum_i \sqrt{r_i s_i/2}$ and    
${\rm Tr}(P_{1}\rho_2)={\rm Tr}(P_{2}\rho_1)=1/2$.  
It is important to note that in general there exist sets of eigenvalues $\{r_i\}$ and 
$\{s_i\}$ 
where $F^2 < 1/4$ and the necessary condition, Eq. (\ref{cond}),
cannot be fulfilled.
In the following, however,  we restrict ourselves to the special case that  
$r_i=s_i$ for $i=1,\dots,d$, for which $F=1/\sqrt{2}$.   
The necessary condition for the lower bound $Q_0$ to be 
achievable then reads $\frac{1}{\sqrt{2}}\leq \sqrt 
{\frac{\eta_2}{\eta_1}} \leq \sqrt{2}$.  
Further, we find the solutions  
$\alpha_{ij}= \alpha \,\delta_{ij}$ and $\beta_{ij}=\beta \,\delta_{ij}$
($i,j=1,\ldots,d$),
where $\alpha=2-\sqrt{\frac{2\eta_2}{\eta_1}}$ and 
$\beta=2-\sqrt{\frac{2\eta_1}{\eta_2}}$. 
$\Pi_0$ has two eigenvalues, $\lambda_{0} = 0$ and $\lambda_{1} = 2-\alpha-\beta$, each with a $d$ fold degeneracy.  Thus the optimum $\Pi_{0}$ is always an operator of rank $d$. 
Note that $2\sqrt{2}-2 \leq \lambda_{1} \leq 1$ in the whole interval 
$F \leq \sqrt{\frac{\eta_{1}}{\eta_{2}}}\leq \frac{1}{F}$. 
Hence in this parameter interval the optimum detection operators
yielding the lower bound  $Q_0$ are 
$\Pi_1=\alpha P_{1\perp}$ and 
$\Pi_2=\beta P_{2\perp}$. 
At the upper and lower limits of the interval the measurement turns into 
the von Neumann measurements that give the failure probabilities 
$Q_{N1\parallel}=Q_{N2}$ and $Q_{N2\parallel}=Q_{N1}$, respectively.   
Since in our example ${\rm Tr}(P_{2}\rho_{1})= {\rm Tr}(P_{1}\rho_{2})=F^2$,
we find that $Q_{N1}= \eta_1 + \eta_2 F^2$ and  
$Q_{N2}= \eta_2 + \eta_1 F^2$. 
Thus we derived a measurement strategy that yields the equality sign 
in Eq. (\ref{rud}) for two mixed states.

In our next examples we focus on the case $d_{1\parallel}
=d_{2\parallel}=1$. 
First we assume that the density operators given in  Eq. (\ref{rho})
have arbitrary ranks $d_1$ and $d_2$, and that    
$\langle r_l|s_m\rangle =  
a \,\delta_{l,1}\delta_{m,1}$ with $|a|<1$.  
This yields $F=\sqrt{s_1 r_1} |a|$, 
${\rm Tr}(P_2\rho_1)=F^2/s_1$,  
${\rm Tr}(P_1\rho_2)=F^2/r_1$ and $d_{1\parallel}=  
d_{2\parallel}= 1$.  
For the parameter range 
specified in Eq. (\ref{cond}) we obtain the optimum detection operators 
\begin{eqnarray}
\Pi_1 =\left(1-\sqrt{\frac{\eta_2}{\eta_1}} \frac{F}{r_1} \right) 
\frac{|\tilde{v}_1\rangle\langle \tilde{v}_1|}{(1-|a|^2)^2} 
+\sum_{l=2}^{d_1}|r_l\rangle \langle r_l|\\  
\Pi_2=\left(1-\sqrt{\frac{\eta_1}{\eta_2}} \frac{F}{s_1} \right) 
\frac{|\tilde{\bar{r}}_1\rangle\langle \tilde{\bar{r}}_1|}{(1-|a|^2)^2} 
+\sum_{m=2}^{d_2}|s_m\rangle \langle s_m|,
\end{eqnarray} 
where 
we introduced 
$|\tilde{v}_1\rangle = |r_1\rangle - a |s_1\rangle$ and 
$|\tilde{\bar{r}}_1\rangle = |s_1\rangle - a^{\ast} |r_1\rangle$.
 
This solution can be applied to the  
problem of {\it quantum state comparison} \cite{jex},  
where two identical quantum objects are each prepared either in the  
state $|\psi_1\rangle$, or in the state $|\psi_2\rangle$,  
and where we wish to determine unambiguously whether the  
states are equal or different.    
The task amounts to distinguishing the  
two-particle states 
$\rho_1= \frac{1}{2}\left( |\psi_1,\psi_1\rangle \langle \psi_1,\psi_1| 
 + |\psi_2,\psi_2\rangle \langle \psi_2,\psi_2|\right)$
and 
$\rho_2= \frac{1}{2}\left( |\psi_1,\psi_2\rangle \langle \psi_1,\psi_2| 
 + |\psi_2,\psi_1\rangle \langle \psi_2,\psi_1|\right)$, 
where $F=|\langle \psi_1|\psi_2\rangle|$. 
Upon determining the eigenstates,  
we find that the structure of $\rho_1$ and $\rho_2$  
corresponds to the one treated in the above special example,  
with $r_1=s_1=(1+F^2)/2$ 
and $|a|= 2F/(1+F^2)$. 
The minimum  
failure probability in unambiguous quantum state comparison 
follows to be 
\begin{equation}
Q_{\rm opt}= \left \{
\begin{array}{ll}
2 \sqrt{\eta_1 \eta_2}F\;\; & \mbox{if 
$\sqrt{\frac{\eta_{\rm min}}{\eta_{\rm max}}} \geq \frac{2F}{1+F^2}$}\\
 \eta_{\rm max} \frac{2F^2}{1+F^2}+
\eta_{\rm min} \frac{1+F^2}{2}\;\; & \mbox{otherwise}.
\end{array}
\right. 
\nonumber
\end{equation}
Here $\eta_{\rm min}$ $(\eta_{\rm max})$ is the smaller (larger) of 
the values $\eta_1=p_1^2+p_2^2$ and $\eta_2=2p_1 p_2$,
where $p_1$ and $p_2$ are the prior probabilities of the states
$|\psi_1\rangle$ and $|\psi_2\rangle$, respectively. 
 
As our final example we mention  
the problem of discriminating a pure state, $\rho_1= |r_1 \rangle \langle r_1|$,   
 from a  mixed state $\rho_2$, or from a set of pure states, respectively, 
that has been introduced as {\it quantum state filtering} \cite{SBH,HB1}. 
In this case ${\rm Tr}(P_1\rho_2)=F^2$ and  
${\rm Tr}(P_{2}\rho_1)=\|r_1^{\parallel}\| ^2$. 
In the parameter interval given by  
Eq. (\ref{cond}) the optimum detection operators take the form  
$\Pi_1=\left(1- \sqrt{\frac{\eta_2}{\eta_1}}F\right) 
\frac{|v_1 \rangle \langle v_1|}{1-\|r_1^{\parallel}\|^2}$ 
  and  
$\Pi_2= \left(1- \sqrt{\frac{\eta_1}{\eta_2}}\frac 
{\|r_1^{\parallel}\|^2}{F}\right) 
\frac{|\bar{r}_1 \rangle \langle \bar{r}_1|}{1-\|r_1^{\parallel}\|^2}+P_2^{\prime}$,   
and the previous solution for the minimum failure probability in  
optimum unambiguous  quantum state filtering \cite{SBH,BHH} is readily regained.

In summary, we performed a detailed analysis of the probabilistic measurement for  
unambiguous discrimination between two arbitrary mixed quantum states.   
We derived general analytical relations that depend on five quantities  
characterizing the   
 mutual relationship of the density operators of the states. 
These quantities are the expressions ${\rm Tr}(P_1\rho_2)$ and  
${\rm Tr}(P_{1\parallel}\rho_2)$, as well as ${\rm Tr}(P_2\rho_1)$ and  
${\rm Tr}(P_{2\parallel}\rho_1)$ and, most importantly, the fidelity $F$.  
 We also showed that the method developed in this paper can be used 
 to find complete analytical solutions that describe the optimum measurement 
 for special cases.

{\it Acknowledgements.} 
We thank  Philippe Raynal and Norbert L\"utkenhaus for sharing a 
manuscript \cite{raynal1} prior to publication on research along very 
similar lines and with partially overlapping conclusions. 
U. H. gratefully acknowledges discussions with O. Benson and the members of his group. 
J. B. acknowledges useful discussions with E.~Feldman, M. Hillery, 
N. L\"utkenhaus, Ph.\ Raynal, and Y. Sun, as well as partial support from the 
Humboldt Foundation and PSC-CUNY.


\begin{thebibliography}{99}   
   
 
\bibitem{springer} see, e. g.,  
J. A Bergou, U. Herzog, and M. Hillery, Lect. Notes Phys. 649, 
417-465 (Springer, Berlin, 2004). 

\bibitem{support}
The support of a density operator is the Hilbert space   
spanned by its eigenvectors with nonzero eigenvalues. The 
rank of the density operor is equal to the dimension of 
the support.  

\bibitem{ivan} I. D. Ivanovic, Phys. Lett. {\bf A123}, 257   
  (1987),    
D. Dieks, Phys. Lett. {\bf A126}, 303 (1988),  
 A. Peres, Phys. Lett. {\bf A128}, 19 (1988).   
   
\bibitem{jaeger} G.\ Jaeger and A.\ Shimony, Phys.\ Lett.\ {\bf A197},   
  83 (1995).     
   
\bibitem{SBH} 
 Y. Sun, J. A. Bergou, and M. Hillery, Phys. Rev. A {\bf      
  66}, 032315 (2002).           
 
\bibitem{BHH}  
J. A Bergou, U. Herzog, and M. Hillery,    
\prl {\bf 90}, 257901 (2003); and \pra {\bf 71}, 042314 (2005).    
 
\bibitem{rudolph} T. Rudolph, R. W. Spekkens, and P. S. Turner,    
  Phys. Rev. A {\bf 68}, 010301(R) (2003).   
   
\bibitem{raynal} Ph. Raynal, N. L\"utkenhaus, and S. van Enk,   
  Phys. Rev. A {\bf 68}, 022308 (2003).   
  
\bibitem{eldar}Y. C. Eldar,  \pra {\bf 67}, 042309 (2003),  
Y. C. Eldar, M. Stojnic, and B. Hassibi, \pra {\bf 69},   
062318 (2004).   
   
\bibitem{feng1} Y. Feng, R. Duan, and M. Ying, \pra {\bf 70},   
012308 (2004).  

\bibitem{HB2} U. Herzog and J. A. Bergou, \pra {\bf 70},   
022302 (2004).    
   
\bibitem{neumark} M.\ A.\ Neumark,    
Izv. Akad. Nauk. SSSR, Ser. Mat. {\bf 4}, 277 (1940).    
   
\bibitem{preskill}  J. Preskill,    
{\it Lecture Notes for Physics 229: Quantum Information and Computation}    
(Cambridge University Press, 1998).   
    
\bibitem{nielsen}  M. A. Nielsen and I. L. Chuang,    
{\it Quantum Computation and Information} (Cambridge University Press, 2000).   
   
\bibitem{footnote} 
Note that $\rho_2|r_l\rangle= \|  r_l^{\parallel}\| \rho_2|h_1\rangle$ 
when  
$d_{1\parallel}=1$. Therefore we can write 
$\sqrt{{\rho}_1}\; {\rho}_2 \sqrt{{\rho}_1} 
= {\rm Tr}(P_{1\parallel}\rho_2){\rm Tr}(P_2 \rho_1) |R\rangle\langle R|$,    
where $|R\rangle=)[{\rm Tr}(P_2 \rho_1)]^{-1/2}\sum_l  
\sqrt{r_l}\|r_l^{\parallel}\| | r_l \rangle$ is a normalized pure state. Hence 
$F=[{\rm Tr}(P_{1\parallel}\rho_2){\rm Tr}(P_2 \rho_1)]^{1/2}$.%
 
\bibitem{jex} S. M. Barnett, A. Chefles, and I. Jex,      
Phys. Lett. A {\bf 307}, 189 (2003).      
 
\bibitem{HB1} U. Herzog and J. A. Bergou, \pra {\bf 65},   
050305 (2002).    

\bibitem{raynal1} Ph. Raynal and N. L\"utkenhaus, quant-ph/0502165.

\end{thebibliography}
\end{document}